\documentclass[aps,pra,twocolumn,amsfonts,floatfix,superscriptaddress,showpacs,nofootinbib]{revtex4}

\pdfoutput=1

\usepackage{graphicx}
\usepackage{epstopdf}
\usepackage{dcolumn}
\usepackage{bm}
\usepackage{mathbbol}
\usepackage{verbatim}
\usepackage{dictsym}
\usepackage{multirow}
\usepackage{ulem}

\usepackage{amssymb}          
\usepackage[latin1]{inputenc} 
\usepackage{amsmath}          
\usepackage{color}
\usepackage{subfigure}
\usepackage{booktabs, tabularx, multirow, array}
\usepackage{units}

\newcommand{\etal}{{\it et al.}}

\newcommand{\ket}[1]{|#1\rangle}

\begin{document}

\title{Heralded single photon sources for QKD applications}

\author{Matteo Schiavon}
\email{matteoschiav@yahoo.it}
\affiliation{Dipartimento di Ingegneria dell'Informazione,
Universit\`a di Padova, via Gradenigo 6/B, 35131 Padova, Italy}
\author{Giuseppe Vallone}
\affiliation{Dipartimento di Ingegneria dell'Informazione,
Universit\`a di Padova, via Gradenigo 6/B, 35131 Padova, Italy}
\author{Francesco Ticozzi}
\affiliation{Dipartimento di Ingegneria dell'Informazione,
Universit\`a di Padova, via Gradenigo 6/B, 35131 Padova, Italy}
\author{Paolo Villoresi}
\affiliation{Dipartimento di Ingegneria dell'Informazione,
Universit\`a di Padova, via Gradenigo 6/B, 35131 Padova, Italy}
\date{\today}

\begin{abstract}
Single photon sources (SPS) are a fundamental building block for optical implementations of quantum information 
protocols.
Among SPSs, multiple crystal heralded single photon sources seem to give the best compromise between high pair production rate and low multiple photon events.
In this work, we  study their performance in a practical quantum key distribution experiment, 
by evaluating the achievable key rates.
The analysis focuses on the two different schemes, symmetric and asymmetric, proposed 
for the practical implementation of heralded single photon sources, with attention on the performance of their composing elements.
The analysis is based on the protocol proposed by Bennett and Brassard in 1984 and on its improvement exploiting decoy state technique.
Finally, a simple way of exploiting the post-selection mechanism for a passive, one decoy state scheme is evaluated.
\end{abstract}

\maketitle

\section{Introduction}
The goal of quantum key distribution (QKD) is to allow two distant parties, Alice and Bob, to share a secret key 
even in the presence of an eavesdropper, Eve.
Since in quantum mechanics measurements irremediably perturb the systems, it is impossible for an eavesdropper to extract useful information without being noticed. Quantum key distribution protocols, like the Bennett-Brassard 1984 (BB84) protocol \cite{Bennett1984}, are proven to be unconditionally secure, when using single photons.
However, due to the spread of laser systems and the difficulty of realizing true single photon sources, 
most QKD implementations use attenuated pulsed lasers as sources.
The need to avoid multi-photon pulses, that can leak information through the photon number splitting (PNS) attack \cite{Brassard2000}, 
requires a low mean photon number per pulse.
This, however, increments the incidence of pulses containing no photons, thus limiting the key generation rate.
The incidence of the PNS attack can be limited using the decoy state technique \cite{Hwang2003,Lo2005,Ma2005}, 
at the expenses of the necessity to modulate the laser intensity.
These limitations have encouraged the research of sources emitting a single photon for each pulse. 

\begin{figure}
	\centering \includegraphics[width=.8\linewidth]{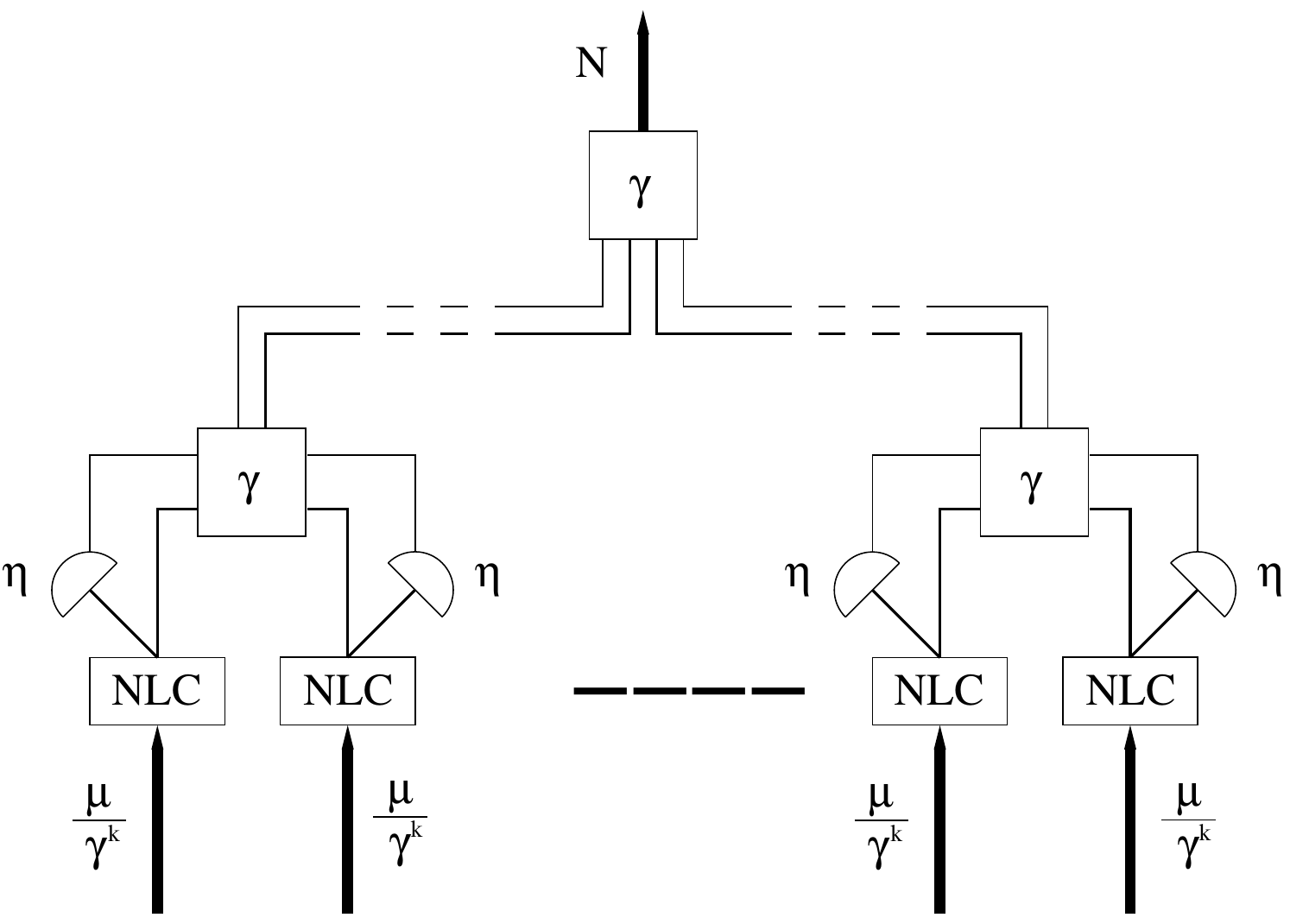}
	\caption{Schematic of the SMHPS \cite{Shapiro2007}. Each non-linear crystal (NLC) is fed with pulses such that the mean number of generated pair per pulse is $\mu/\gamma^k$, with $k = \log_2 m$ and $\gamma$ is the transmittance of 2-to-1 optical switches. The idler of each NLC is fed into a detector with quantum efficiency $\eta$.}
	\label{fig:SMHPS}
\end{figure}

One possible implementation of these sources is based on localized quantum structures, 
such as color centres \cite{Aharonovich2011}, quantum dots \cite{Pelton2002,Ayesha2014}, atoms \cite{McKeever2004}
 or ions \cite{Keller2004} in a cavity.
These schemes, however, show some major drawbacks: 
the need of expensive equipment for their operation and limitations in wavelength and bandwidth selection \cite{Shapiro2007}.
An alternative scheme for single photon sources exploits the process of spontaneous parametric down conversion (SPDC) in a non-linear crystal.
An intense laser pump on a non-linear crystal leads to the probabilistic emission of pairs of photons, called signal and idler.
The idler photon can be used to ``herald'' the presence of the signal photon, giving the so called heralded source (HS).
If the duration of the pulse is much greater than the reciprocal of the phase-matching bandwidth
the statistics of the pairs is still poissonian \cite{Shapiro2007}.
There are some strategies for improving the single-photon character of the heralded source.
One strategy consists in using a single HS with photon number resolving detector on the idler channel and selecting 
the pulses where only one photon has been detected \cite{Lasota2013}.
An alternative strategy uses parallel HS units and post-selection: 
each unit is pumped with low intensity to suppress multi-photon events, while keeping the overall rate at an acceptable level.
The scheme of multiple HS with post-selection (MHPS) has been originally proposed by Migdall \etal~\cite{Migdall2002}:
it used an $m$-to-1 optical switch triggered by a detector on the idler photon of each HS.
Migdall \etal~took into account also the finite efficiency of single photon detectors but, 
as pointed out by Shapiro and Wong \cite{Shapiro2007}, the current low efficiency of $m$-to-1 
optical switches strongly limits the performance of this architecture.
To overcome this limitation, they proposed a symmetric scheme (SMHPS) using $m$ HS units linked by $m-1$ 
binary polarization-based photon switches in a tree structure \cite{Shapiro2007}, see Fig. \ref{fig:SMHPS}.
They studied the probability of single-photon and multi-photon pulses, with real detectors and optical switches.
Their scheme has been subsequently implemented using 4 crystals, both with bulk optics \cite{Ma2011} and using hybrid photonic circuits \cite{Meany2014}. 

\begin{figure}
	\centering \includegraphics[width=.8\linewidth]{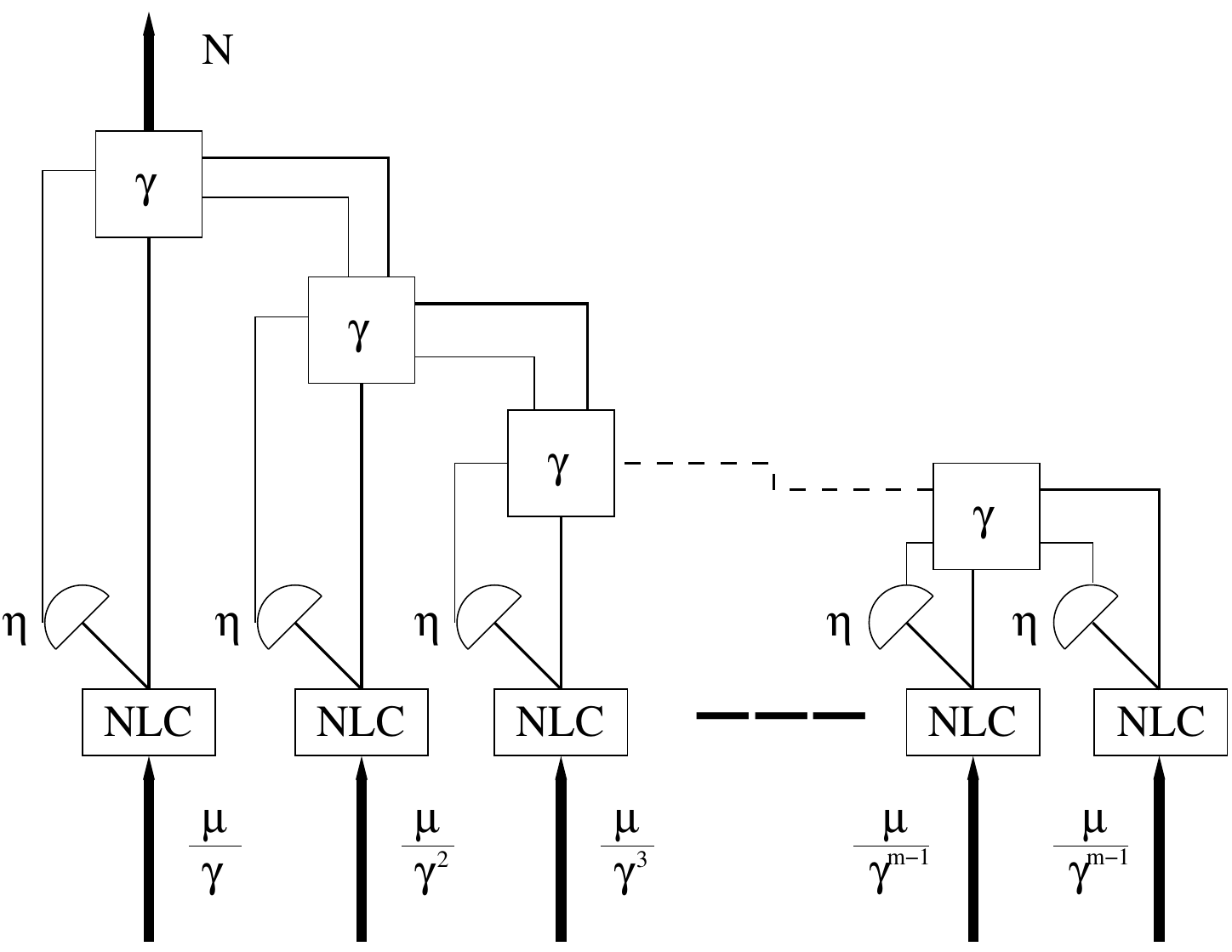}
	\caption{Schematics of the AMHPS \cite{Mazzarella2013}. Each non-linear crystal (NLC) is pumped with a different intensity in order to compensate the different number of traversed 2-to-1 optical switches, each characterized by its transmittance $\gamma$. The idler of each NLC is fed into a detector with quantum efficiency $\eta$.}
	\label{fig:AMHPS}
\end{figure}

A thorough analysis of multiple heralded sources with post-selection has been performed in \cite{Mazzarella2013}.
From their work, it emerges that, in the case of imperfect devices, the symmetric scheme suffers from a scalability issue, 
with a decrease in one photon probability when increasing the number of crystals.
In \cite{Mazzarella2013}, a new asymmetric scheme (AMHPS) that does not present this problem
was also proposed.
The scheme is shown in Fig. \ref{fig:AMHPS} for completeness.
The SMHPS and AMHPS schemes were also referred respectively as {\it log-tree} and {\it chained scheme} in \cite{bonn2015njp}.

Sub-poissonian sources have already been demonstrated to give an improvement over poissonian ones in 
the key rate of a QKD system \cite{Waks2002,Lasota2013}.
This article specializes the analysis to the case of the multiple-crystal heralded photon sources studied in \cite{Mazzarella2013}, 
parametrizing them using their design parameters instead of the more experimentally relevant efficiency and second-order correlation.
Moreover, the security analysis is extended to the case of general attacks by Eve \cite{Scarani2008}.

In this work, we follow \cite{Mazzarella2013}, 
and compare the performances of the different architectures and their efficacy in a practical QKD implementation.
An ideal MHPS in the configuration proposed by Migdall \etal, with perfect heralding efficiency and an $m$-to-1 optical switch with no losses, is studied first, in order to give an account of the maximum performance that can be obtained with this kind of sources.
Next, two architectures for implementing a multiplexed source in a realistic scenario are analyzed, based on {\it symmetric} and {\it asymmetric} networks of 2-to-1 switches. Finite heralding efficiency and optical switch transmittance are considered in order to illustrate the potential of these source with state-of-the-art technology.
The different types of sources are inserted in a common model of QKD, based on the BB84 protocol and taking into account channel and detector inefficiencies \cite{Scarani2008}.
The optimization maximizes the key generation rate for the different architectures, with different numbers of HS units, for a wide range of channel losses.
The obtained values are then compared with the key generation rate of both an ideal single photon source and an attenuated laser. 

The symmetric and asymmetric schemes are also optimized when used with decoy states \cite{Hwang2003,Lo2005}, both in an active \cite{Ma2005} and in a passive scheme \cite{Adachi2007,Curty2009}.
The active scheme, where Alice actively changes the output statistics of its apparatus, allows her to choose an arbitrary number of decoy states, giving her the ability to estimate channel parameters with arbitrary precision.
Passive decoy, on the other hand, is implemented by exploiting photon number correlations of the two outputs of non-linear crystals \cite{Adachi2007}.
This gives less flexibility in the choice of the decoy states, thus limiting the precision of parameter estimation, but allows to reach a key generation rate comparable with active decoy without introducing further complexity in Alice's apparatus.


\section{Model}
\label{sec:model}
To evaluate the performances of the heralded single photon sources, we consider the 
Bennett-Brassard 1984 (BB84) QKD protocol \cite{Bennett1984} with polarization encoding.
Alice, the transmitter, encodes the random bits into the $Z\equiv\{\ket{H},\ket{V}\}$ or the $X\equiv\{\ket{D},\ket{A}\}$ basis, where $H$, $V$, $D$ and $A$ refer to the horizontal, vertical, diagonal and anti-diagonal polarization respectively.
Bob randomly measures the incoming photon into the $Z$ or $X$ basis.
After the transmission, Alice and Bob discard the events in which the sent and measured basis are different. 
On the remaining events, they evaluate the secret key rate, namely the amount of unconditional secret bits that they have produced.

The single photon source held by Alice is completely described by its output statistics. We indicate the probability
of emitting $n$ photon by
$P_n^{M}(\mu; m)$ for the MHPS and $P_n^{S}(\mu; m, \eta, \gamma)$ and $P_n^{A}(\mu; m, \eta, \gamma)$ for the SMHPS and the AMHPS,
respectively \cite{Mazzarella2013} (see Appendix \ref{sec:source} for a more detailed description of the different architectures and the explicit expressions of $P_n^{M}$, $P_n^{S}$ and $P_n^{A}$) .
The source is pumped by a pulsed laser, with phase randomization between each pulse \cite{Scarani2008}.
The variable $\mu$ is proportional to the intensity of the pump laser, which is related to the mean number of pairs generated by each HS per pulse.
The sources are parametrized by the number of non-linear crystals (i.e. of HS units) $m$ and, for the SMHPS and the AMHPS, by the efficiency of the heralding detectors $\eta$ and the transmittance of the optical switches $\gamma$.
For $\gamma \to 1$, the scheme used in the switching network is no longer significant, therefore the output statistics of the two architectures coincide.
If the heralding efficiency $\eta \to 1$ as well, the two architectures become equivalent to the ideal MHPS.

We used as model for the QKD channel a depolarizing lossy channel (DLC), characterized by a transmittance $t = 10^{-\frac{L}{10}}$, with $L$ the loss level in decibel, and a depolarization effect with visibility $V$, that takes into account also alignment and stability issues. 

The receiver consists of an optical apparatus, characterized by transmittance $t_B$, and two single photon detectors, each with quantum efficiency $\eta_B$ and dark count probability $p_d$.
They are threshold detectors, i.e. they cannot discriminate the number of incident photons.
Assuming the effects of the channel on each photon of an $n$-photon pulse are independent, 
the probability that a detector clicks, when an $n$-photon signal is sent, is
\begin{equation}
	\eta_n = 1 - (1 - \eta_D t_B t)^n.
\end{equation}  

The parameters used for the evaluation of the secret key rate are the \textit{gain} $Q$, defined as the probability that a pulse gives a click in Bob's measurement apparatus, and the \textit{quantum bit error rate} (QBER) $E$, i.e. the error probability in Bob's detection events.
These are the only measurable parameters during a QKD experiment.
For the DLC they can be estimated as follow (see also \cite{Ma2005}). 

The gain is defined as
\begin{equation}
	Q = \sum_{n=0}^\infty Y_n P_n,
\end{equation}
where $P_n$ is the probability of having $n$ photons in a pulse and $Y_n$ is the yield of an $n$-photon signal, i.e. the conditional probability of a detection event at Bob's side given that Alice sends $n$ photons.
Assuming independence between signal and background, the yield of an $n$-photon pulse for a DLC can be predicted to be
\begin{equation}
	\label{eq:Y_n}
	\widetilde Y_n = \widetilde Y_0 + \eta_n - \widetilde Y_0 \eta_n \simeq \widetilde Y_0 + \eta_n,
\end{equation}
where $Y_0$ is the probability of a dark count event, which is $\widetilde Y_0 \simeq 2 p_d$ in the case of two independent detectors and small dark count probability.
From now on we use the convention of indicating with tilde the predicted parameters for a DLC.
The negative term, coming from the fact that real detection events and dark counts are not mutually exclusive, can be neglected, since $Y_0 \ll 1$.  

The QBER is defined by
\begin{equation}
	E = \frac{1}{Q }\sum_{n=0}^\infty e_n Y_n P_n,
\end{equation}
where $e_n$ is the $n$-photon error rate, i.e. the probability of an error when Alice sends a $n$-photon state.
For a DLC the $n$-photon error rate can be predicted to be
\begin{equation}
	\label{eq:e_n}
	\widetilde e_n = \frac{\widetilde e_0 \widetilde Y_0 + \widetilde e_d \eta_n}{\widetilde Y_n},
\end{equation}
where $\widetilde e_0 = \frac{1}{2}$ is the error probability of a dark count event, which is assumed to be random, and 
$\widetilde e_d = \frac{1-V}{2}$ is the probability that a photon hits the wrong detector.

\section{Secret key rate}
\label{sec:rate}
After the transmission, Alice and Bob use post-processing to extract a shared, secret key from the exchanged symbols.
The secret key rate $R$ is defined as the fraction of pulses that produce a secret bit (without counting those discarded in the sifting phase) \cite{Scarani2008}.
Its high dependence on source statistics makes it the most suitable parameter for the comparison of the different configurations.

\subsection{BB84 without decoy state}
The rate of the BB84 protocol is limited by the fact that Eve can, in principle, obtain full information from a multi-photon pulse through the photon number splitting (PNS) attack, without introducing any error \cite{Brassard2000}.
In the asymptotic limit of infinite key, the achievable key rate is
\begin{equation}
\label{nodecoy}
	R = Q \{ (1-\Delta) [ 1 - h\left( \frac{E}{1-\Delta} \right) ] - f_{EC} h(E) \},
\end{equation}
where $\Delta$ is the multi-photon rate, defined as
\begin{equation}
	\Delta = \frac{1 - P_0 - P_1}{Q},
\end{equation}
$f_{EC}$ is the error correction efficiency and $h(x)$ is the binary Shannon entropy \cite{Scarani2008,Gottesman2002}.
For true single photon sources $\Delta=0$ and the secret key rate is written as $R = Q [ 1 - h(E)  - f_{EC} h(E) ]$: the correction term $1-\Delta$ in \eqref{nodecoy}, indeed, takes into account the possible PNS attack on the multi-photon pulses.

\subsection{BB84 with active decoy}
The decoy state technique has been introduced to counteract the PNS attack \cite{Lo2005}.
It consists on randomly varying the source statistic, so that Eve can no longer adapt her attack to Alice's state.
After the transmission, Alice communicates Bob the state she used for every pulse, allowing them to estimate channel parameters conditioned to that knowledge. 

The decoy state technique is active in the sense that Alice chooses the output statistics using a random number generator and (typically) a variable attenuator after the source.
In principle, she can choose an arbitrary number of decoy states: however it has been shown that just using the vacuum and a weak decoy state gives tight bounds on the relevant parameters \cite{Ma2005}.
In the asymptotic limit of infinite key, the key rate is
\begin{equation}
	\label{eq:active_rate}
	R = P_0 Y_0 + P_1 Y_1 [ 1 - h(e_1) ] - Q f_{EC} h(E),
\end{equation}
where $P_0$ and $P_1$ are given by the source statistics in the signal state and the parameters $e_1$, $Y_0$ and $Y_1$ are the channel parameters estimated using decoy states \cite{Ma2005,Lo2005-1}.
Following \cite{Scarani2008}, we make the simplifying assumption that the parameters have been determined exactly.

\subsection{BB84 with passive decoy}
In passive decoy state QKD, the source statistics is not under Alice's direct control, but is conditioned on some random event at Alice's side.
A typical example consists in an attenuate coherent state passing through a 50/50 beam splitter (BS) with a single photon detector at the reflected output: the photon statistic at the transmitting output of the BS changes when Alice detects or not a photon.
In the case of heralded sources, we denote by $P_n^{(c)}$ or $P_n^{(nc)}$ the output statistics if, respectively, at least one detector or no detector clicks (see Appendix \ref{sec:passive} for the explicit form of these probabilities for the different architectures).
We note that $P_n^{(nc)}$ is not trivial since in both schemes the post-selection mechanism outputs 
the first HS if no detector fires \cite{Mazzarella2013}.
This feature can be used to implement a passive decoy state, since Eve has no way of distinguishing the statistics of each pulse before it is publicly announced.

Assuming the post-processing is done separately for each statistics, the key rate is
\begin{equation}
	\label{eq:passive_rate}
	R = P^{c} R^{c} + P^{nc} R^{nc},
\end{equation}
where $R^{c}$ and $R^{nc}$ are the key rate for, respectively, the case of at least one detector and no detector clicking.
The key rate is, in the limit of infinite key,
\begin{equation}
	\label{eq:passive_partial_rate}
	R^{\xi} = P_0^{(\xi)} Y_0^L + P_1^{(\xi)} Y_1^L [ 1 - h(e_1^U) ] - Q^{\xi} f_{EC} h(E^{\xi}),
\end{equation}
where $\xi \in \{ c, nc \}$, $Q^{\xi}$ and $E^{\xi}$ are the parameters estimated from the pulses in the corresponding statistics and $Y_0^L$, $Y_1^L$, $e_1^U$ are the lower (L) and upper (U) bounds for the parameters estimated from $\{Q^{c}, E^{c}, Q^{nc}, E^{nc}\}$ and the known source statistics.
The explicit formulas for parameter estimation, derived from \cite{Curty2010}, are given in Appendix \ref{sec:passive} in eqs. \eqref{Y0L}, \eqref{Y1L} and \eqref{e1U}.

\section{Results}
In the present section we compare the performances of the SMHPS and AMHPS sources
for different values of $m$, namely the number of HS units considered.
The parameter $\mu$, related to the number of generated pairs per pulse, is the free parameter used to
numerically maximize the rate.

For the channel and detector parameters, we used typical values of present day fibre-based QKD systems \cite{Scarani2008}.
We consider a channel with visibility $V = 0.99$ and losses ranging from $0$ to $\unit[55]{dB}$ 
(we note that $\unit[55]{dB}$ corresponds to 
$\unit[275]{km}$ if we consider the typical fibre attenuation of $\alpha = \unit[0.2]{dB/km}$).
Bob's apparatus is characterized by optical transmittance $t_B = 1$ and detectors with quantum efficiency $\eta_B = 0.25$ and dark count probability $p_d = 2 \cdot 10^{-7}$, corresponding to the state-of-the-art of infrared semiconductor single photon detectors \cite{id230}. 
The efficiency of the error correction code is $f_{EC} = 1.05$ \cite{Mateo2015}. 

All the simulated key rates are compared with the one obtained with an attenuated laser source (WCS) of poissonian output statistics $P_n = e^{-\mu} \mu^n/n! $, where $\mu$ is the mean number of photons per pulse, both without and with decoy state.
In the passive scheme, the comparison is extended to the one decoy state with attenuated laser described in \cite{Ma2005}, where also the inefficiencies in parameter estimation are taken into account.
Furthermore, all schemes are compared with the single photon case, representing an upper bound of the 
rate attainable in a given configuration.

In the following subsections the performances of the heralded single photon source are compared in different cases.

\subsection{Performances of BB84 without decoy state}
\begin{figure}[t]
	\centering \includegraphics[width=0.9\linewidth]{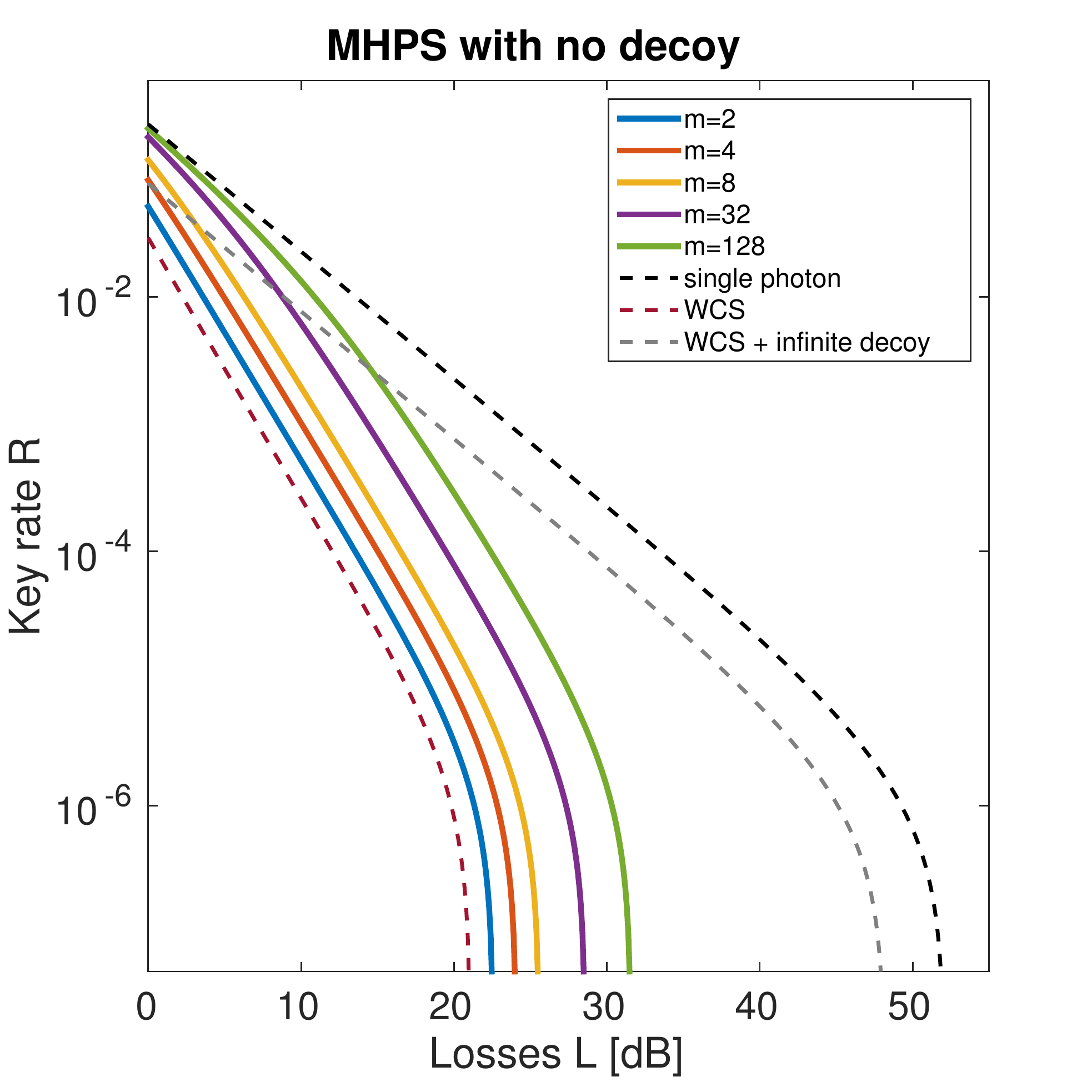}
	\caption{Key rate for the MHPS architecture in function of channel losses.}
	\label{fig:single}
\end{figure}

The ideal MHPS gives the maximum key rate attainable with this kind of sources \cite{Mazzarella2013}.
This value is shown in Figure \ref{fig:single}.
The key rate increases with the number of HS units and approaches the single photon case for $m=128$, in the low loss regime.
Indeed, for $m \to \infty$, $\Delta = O(\mu)$ and $Q \simeq 1$, approximating the single photon case for $\mu \ll 1$.
At increasing losses, however, the contribution of multi-photon pulses increases and the source shows the same behaviour as the attenuated laser.
This had already been observed in \cite{Waks2002}, with the difference that, for low $m$, not only the fraction of multi-photon pulses is higher, determining the lower maximum tolerable loss level, but also the incidence of pulses with zero photons is stronger, determining the lower key rate at $L = \unit[0]{dB}$. 

\begin{figure*}[t]
	{\includegraphics[width=0.45\linewidth]{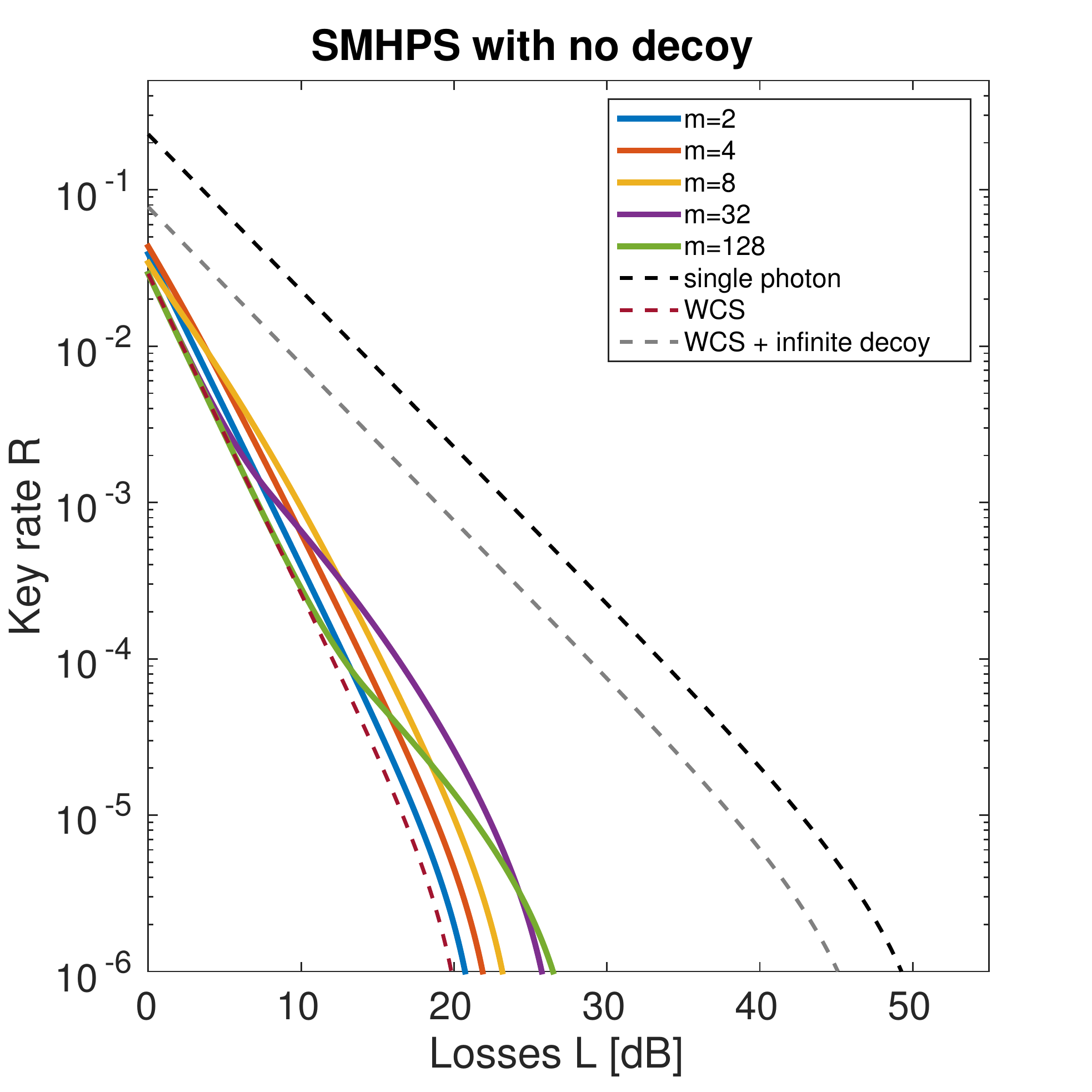}}
	{\includegraphics[width=0.45\linewidth]{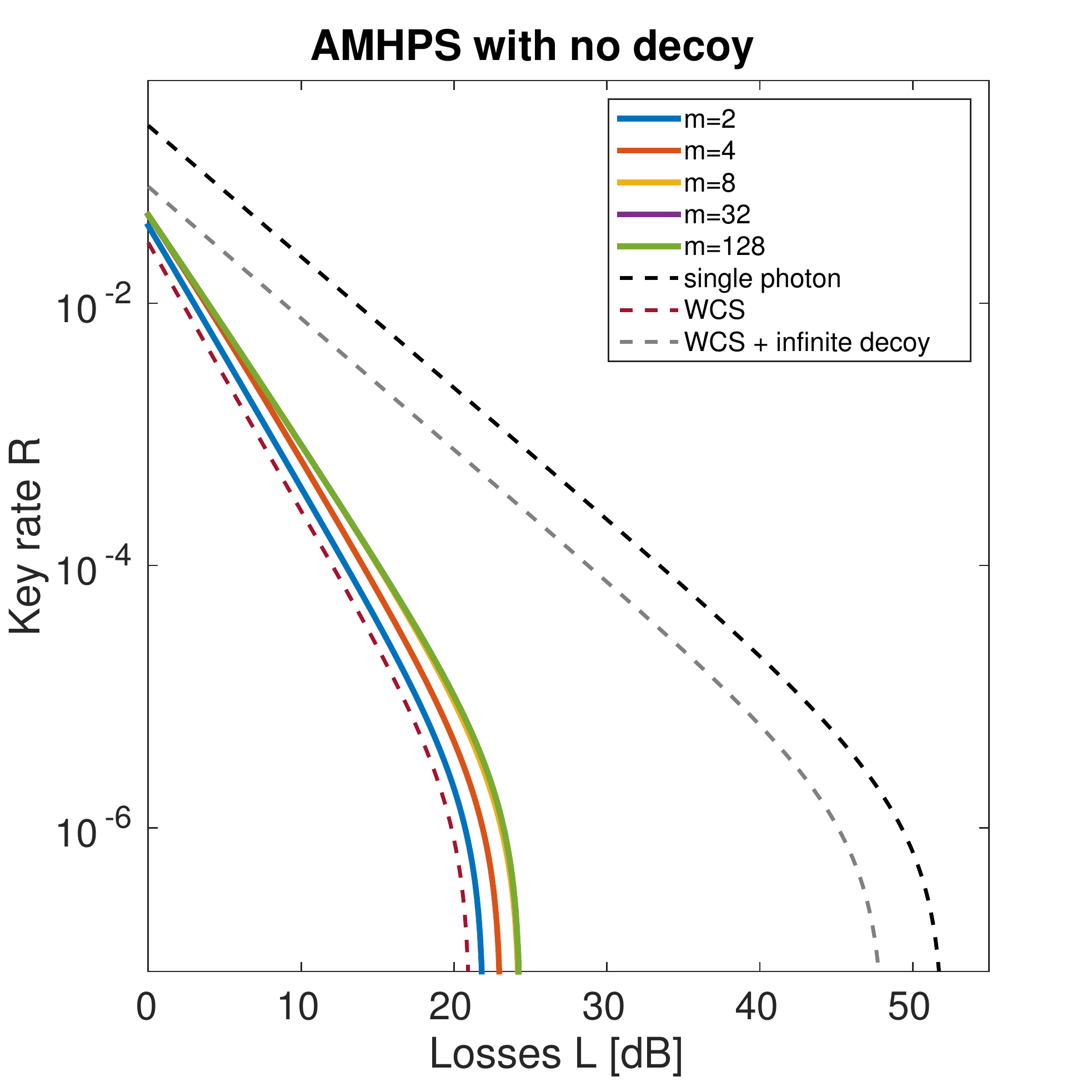}}
	\caption{Key rate of the SMHPS (left) and the AMHPS (right), with $\eta = 0.7$ and $\gamma = 0.5$ for BB84 without decoy state. For the AMHPS (right), the curves for $m=8$, $m=32$ and $m=128$ are superposed.}
	\label{fig:symm_asymm}
\end{figure*}

While these limitations are proper of the multiple-crystal architecture itself, the implementation using finite efficiency devices further reduces the key generation rate.
Using $\eta = 0.7$ and $\gamma = 0.5$ as, respectively, heralding efficiency and switch transmittance \cite{Mazzarella2013}, the key generation rate of the SMHPS and the AMHPS has the values shown in Figure \ref{fig:symm_asymm}.
Both the key rate and the maximum tolerable losses are lower than for the ideal MHPS, since the low switch transmittance requires a higher mean number of generated pairs per pulse in order to avoid zero-photon pulses, thus increasing also the incidence of multi-photon pulses.
This effect is particularly evident in the case of the SMHPS, where, independently from the HS unit triggered to output, the number of switches crossed scales as $k = \log_2 m$.
For few HS units, where the number of crossed switches is low, the predominant effect is the suppression of multi-photon events, therefore the key rate increases with $m$.
For a higher number of HS units (such as $m=32$ or $m=128$), the source shows a low key rate for low losses (not much higher than the attenuated laser one) and a high maximum tolerable loss level.
In the limit $m \to \infty$, any advantage over the attenuated laser is lost.

On the other hand, the AMHPS has a more stable behaviour, since its key rate never decreases by increasing $m$, but reaches an optimal value and then remains unchanged (the key rates for $m = 8$ and $m=128$ are almost equal).
This is due to the fact that, when a certain number of HS units has been reached, the addition of further HS units does not yield significant improvement, since the probability that the rightmost HS units are triggered to the output is negligible and the output is given only by the leftmost HS units, whose configuration does not change (see Figure \ref{fig:AMHPS}). 

\begin{figure*}[t]
	{\includegraphics[width=0.45\linewidth]{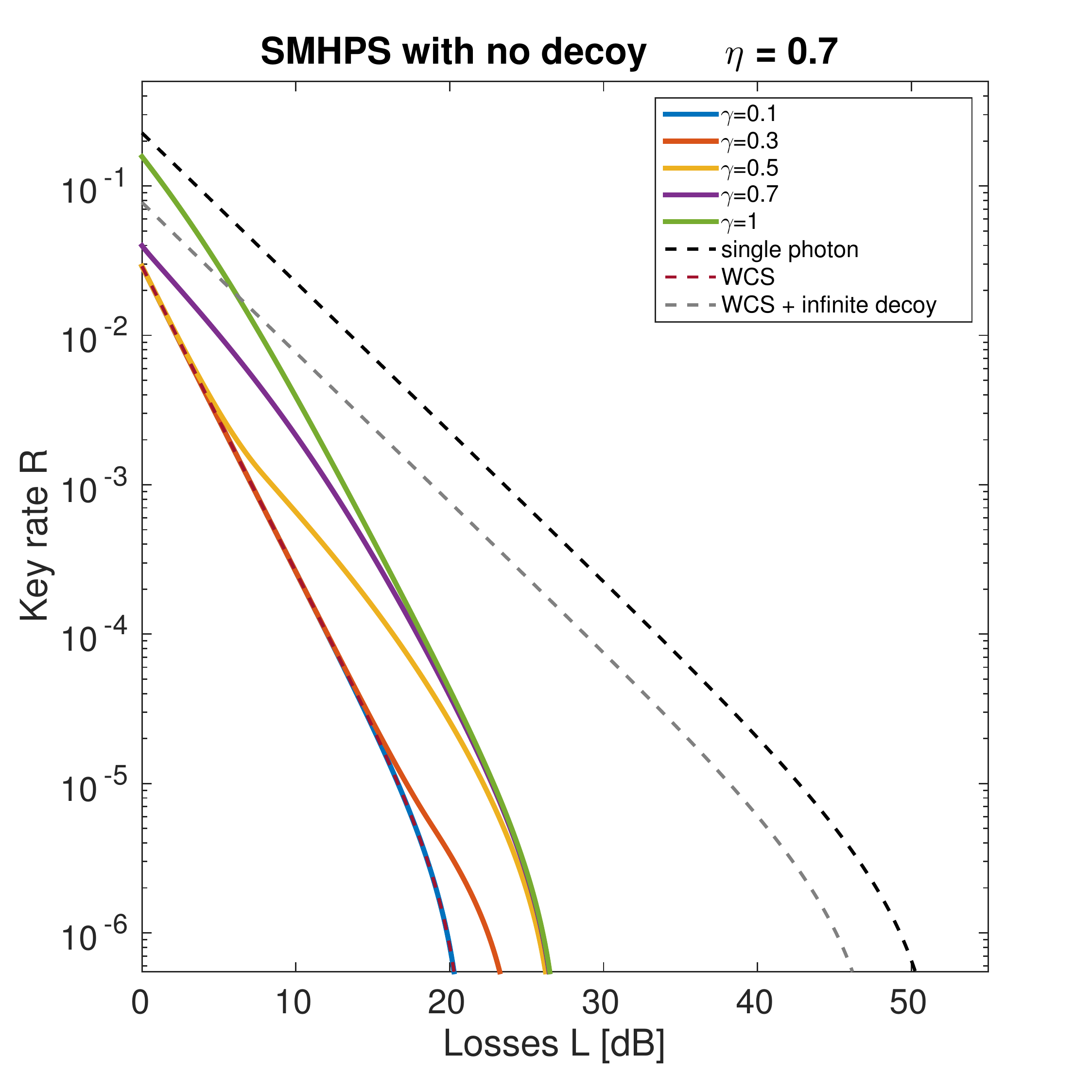}}
	{\includegraphics[width=0.45\linewidth]{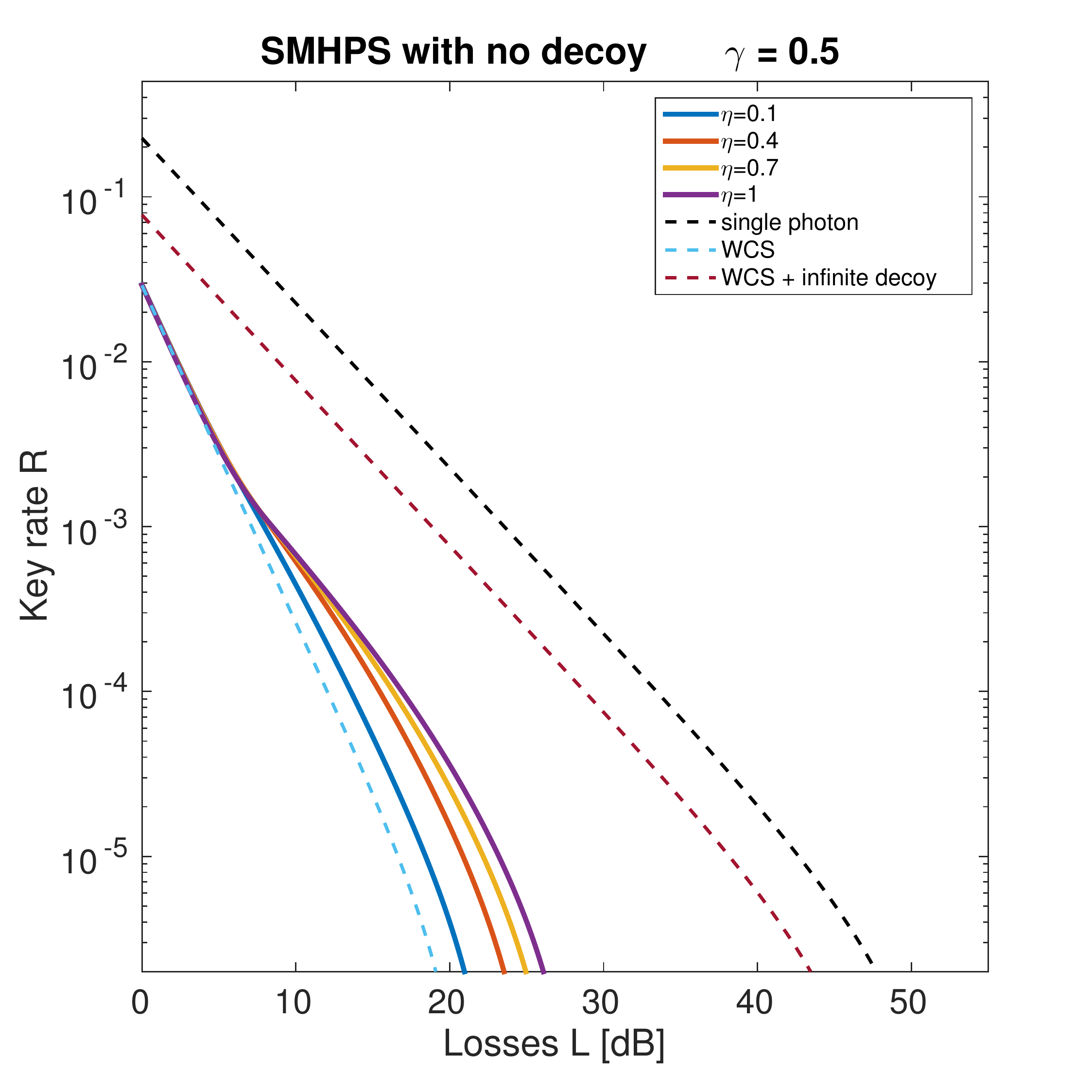}}
	\caption{Key rate (without decoy state) for the SMHPS for $m=32$ and (left) $\eta = 0.7$ and different values of $\gamma$, (right) $\gamma = 0.5$ and different values of $\eta$.}
	\label{fig:symm_hg}
\end{figure*}

\begin{figure*}%
	{\includegraphics[width=0.45\linewidth]{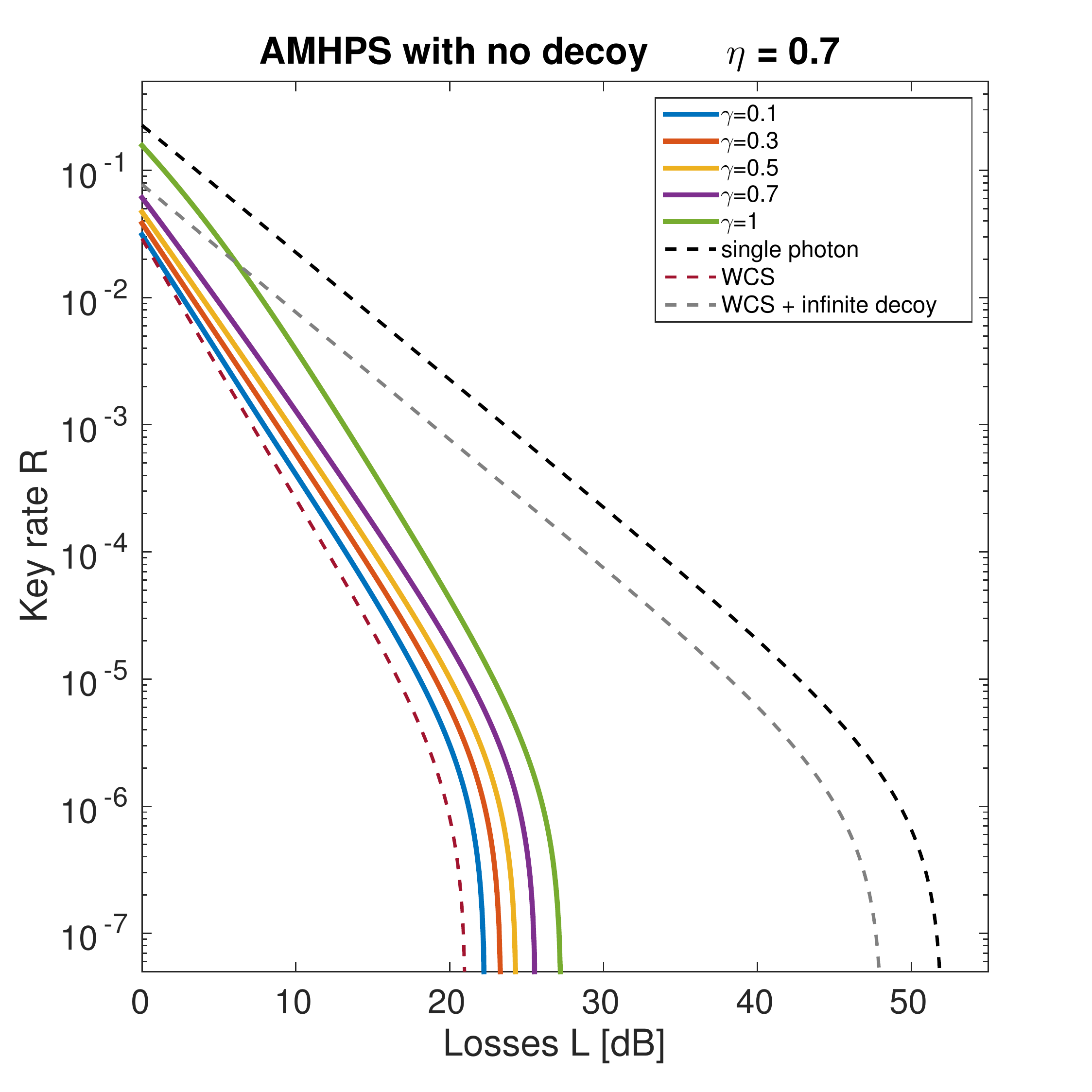}}
	{\includegraphics[width=0.45\linewidth]{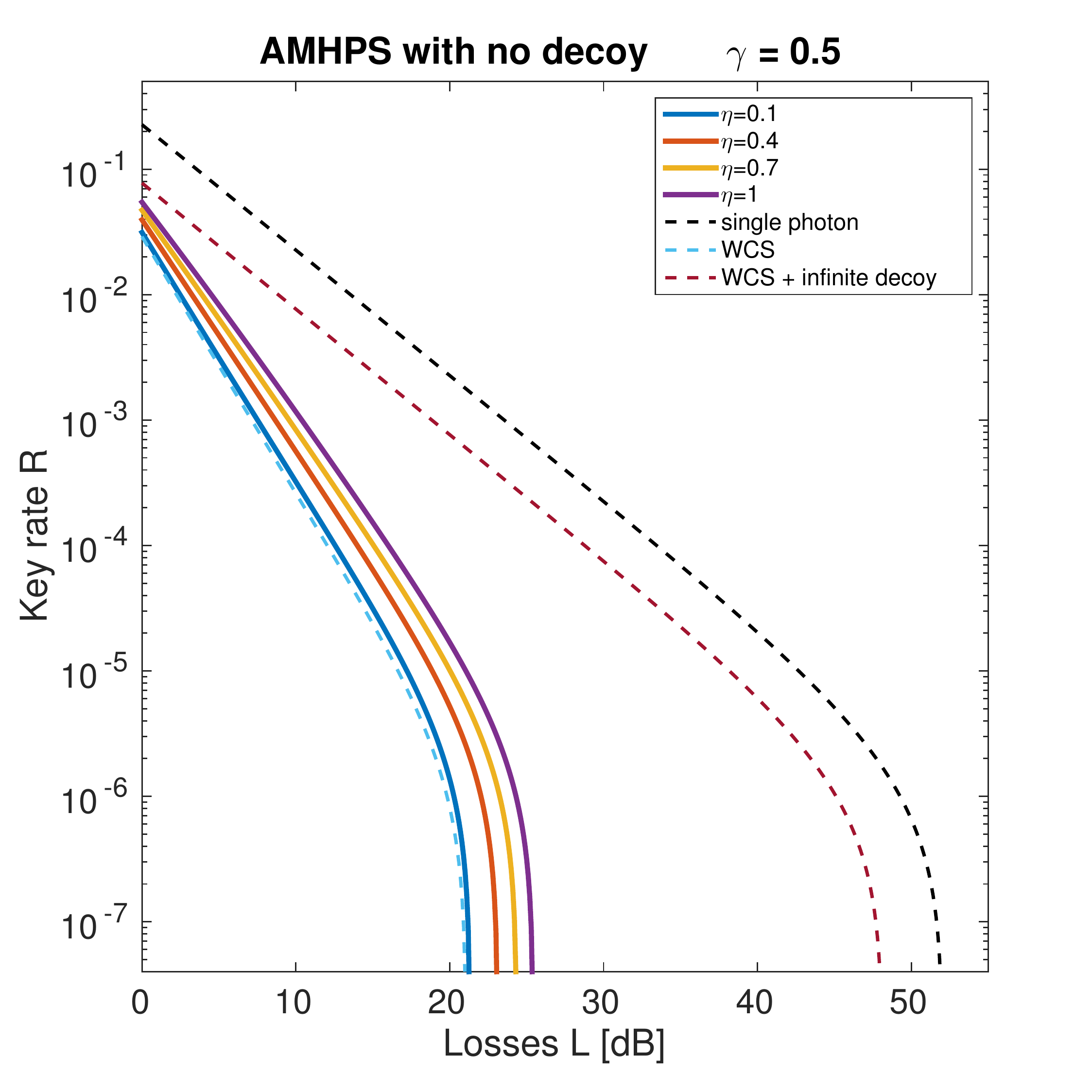}}
	\caption{Key rate (without decoy state) for the AMHPS for $m=32$ and (left) $\eta = 0.7$ and different values of $\gamma$, (right) $\gamma = 0.5$ and different values of $\eta$.}
	\label{fig:asymm_hg}
\end{figure*}

The different behaviour of the two architectures is even more evident when studying their key rate at the variation of the two relevant source parameters, the detection efficiency $\eta$ and the switch transmittance $\gamma$.
The key rate of the SMHPS and the AMHPS is shown, respectively, in Figures \ref{fig:symm_hg} and \ref{fig:asymm_hg}, when fixing $\eta = 0.7$ and changing $\gamma$ on the left and with fixed $\gamma = 0.5$ and changing $\eta$ on the right.
The behaviour of the SMHPS is highly dependent on the switch transmittance.
For low $\gamma$, the benefits deriving from multiple HS units do not compensate the higher absorption rate.
As evident from Figure \ref{fig:symm_hg}, for $\gamma < 0.5$ the SMHPS does not perform much better than the attenuated laser.
This is consistent with the results from \cite{Mazzarella2013}, where it has been shown that, in the asymptotic limit $m \to \infty$, the SMHPS performs better than the attenuated laser for $\gamma \geq 0.5$.
The curve $\gamma = 0.5$ corresponds to the transition  between the laser-like regime and the MHPS-like one. 

The effect of the detector efficiency $\eta$ is evident in the high loss regime, where the influence of the lower number of multi-photon pulses is more important.
This can be evinced from the right plot of Figure \ref{fig:symm_hg}, where the detection efficiency $\eta$ is varied for the case $\gamma = 0.5$.
In the low loss regime, the key rate is the same for all values of $\eta$, showing that in this region the dominating effect is photon absorption in the optical routing.
On the other hand, the high loss regime shows an improvement in the maximum tolerable loss level from  $\unit[21]{dB}$ for $\eta = 0.1$ to  $\unit[26]{dB}$ for $\eta = 1$.
The increased detection efficiency allows a better choice of the HS unit to route to output, thus allowing the HS units to be pumped with less intensity and decreasing the incidence of multi-photon pulses. 

The key rate curves of the AMHPS, on the contrary, show the same trend for all tested combinations $(\eta,\gamma)$.
In the asymmetric scheme, indeed, photons emitted by the leftmost HS units pass a low number of optical switches before being routed to output, therefore the effect of photon absorption never dominates over multi-photon pulses.

\subsection{Performances of BB84 with decoy state}
\begin{figure*}[t]
	{\includegraphics[width=0.45\linewidth]{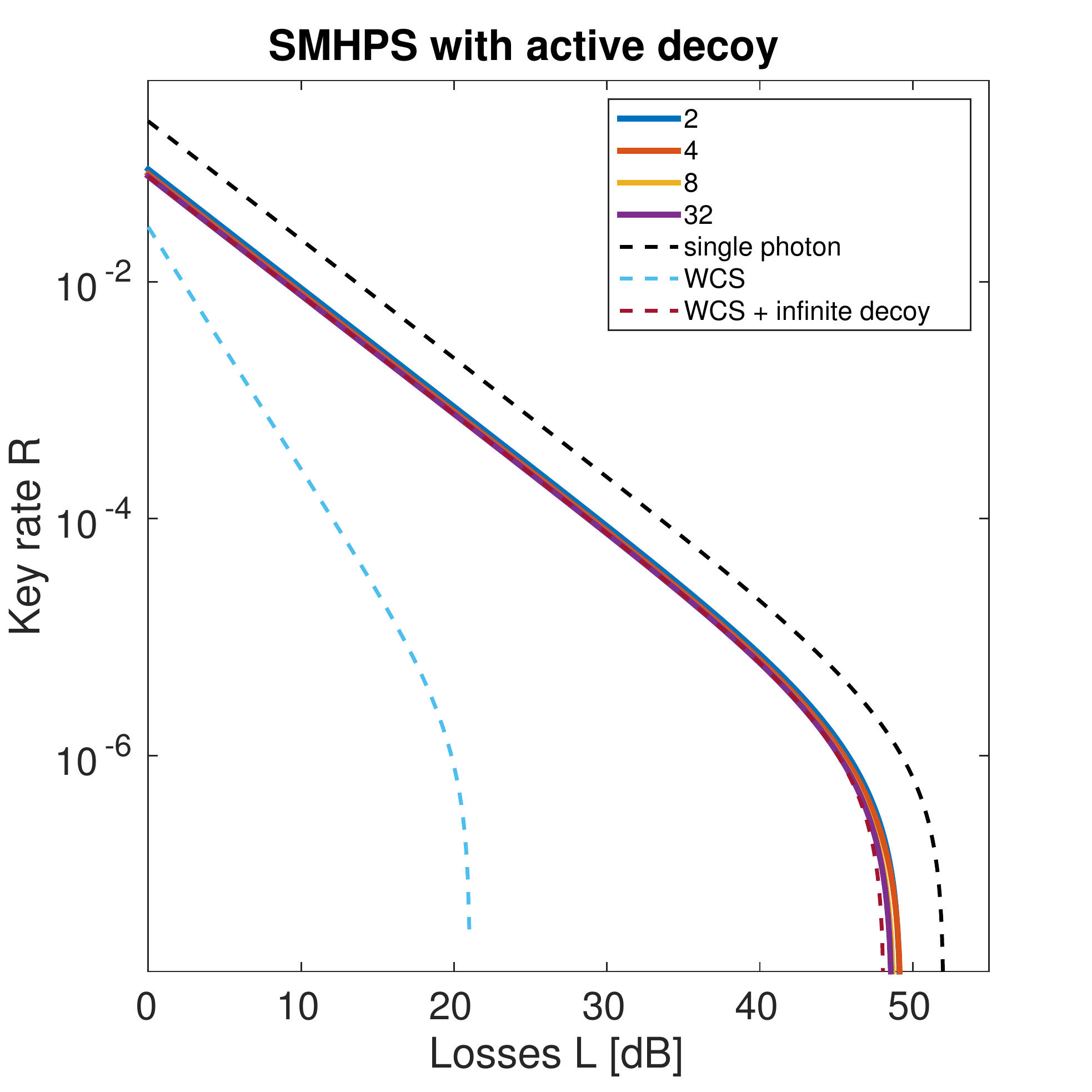}}
	{\includegraphics[width=0.45\linewidth]{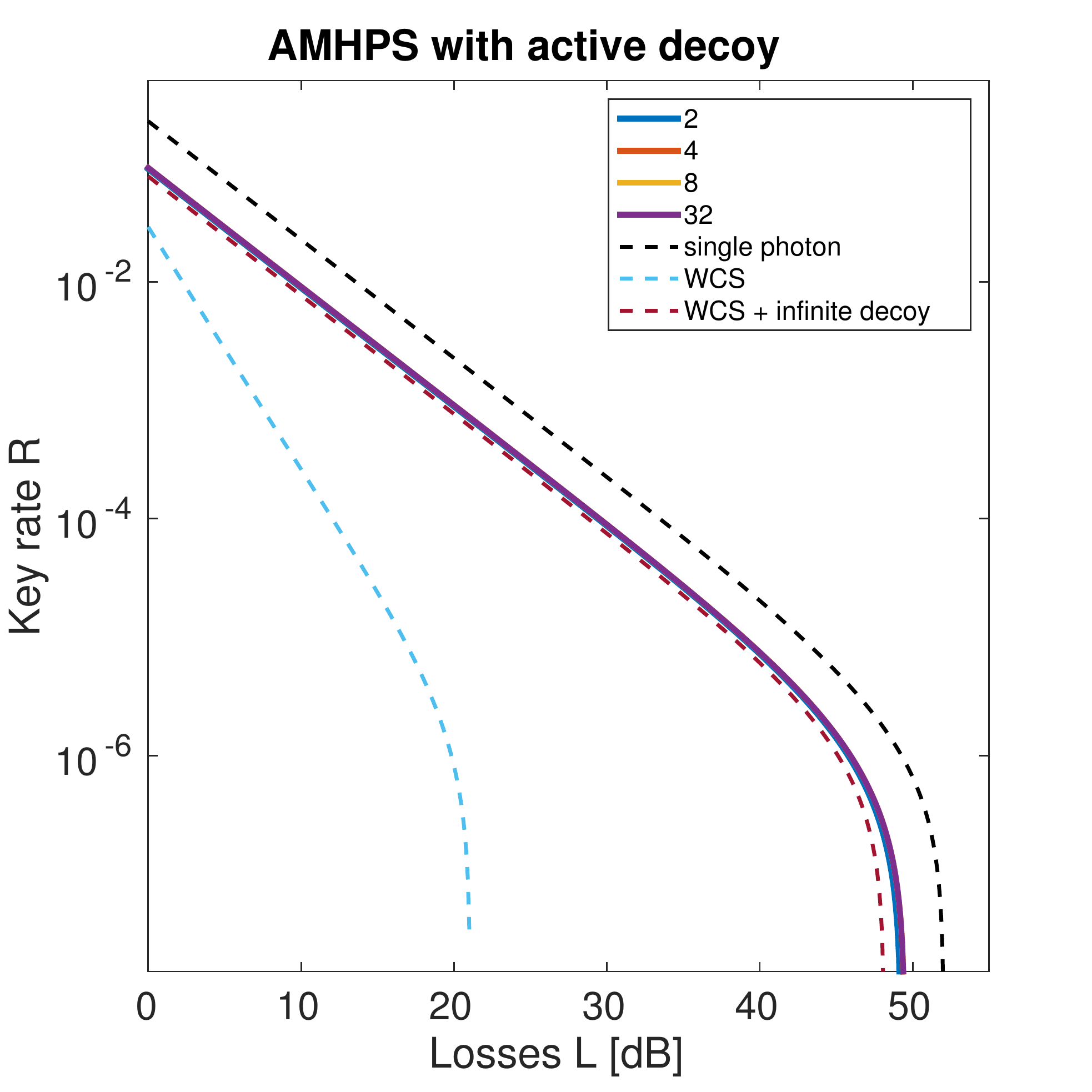}}
	\caption{Key rate of active decoy state QKD for the SMHPS (left) and the AMHPS (right), with $\eta = 0.7$ and $\gamma = 0.5$. For the SMHPS (left), the curves are very close, with $m=32$ the lowest curve. In the AMHPS case (right), the lowest curve has $m=2$, while all the others are superposed.}
	\label{fig:active}
\end{figure*}

The key rate for active decoy is obtained by using equation (\ref{eq:active_rate}), with $e_1$, $Y_0$ and $Y_1$ calculated by using the channel parameters given in (\ref{eq:e_n}) and (\ref{eq:Y_n}).
The results of simulations are shown in Figure \ref{fig:active}.
\begin{figure*}[t]
	{\includegraphics[width=0.45\linewidth]{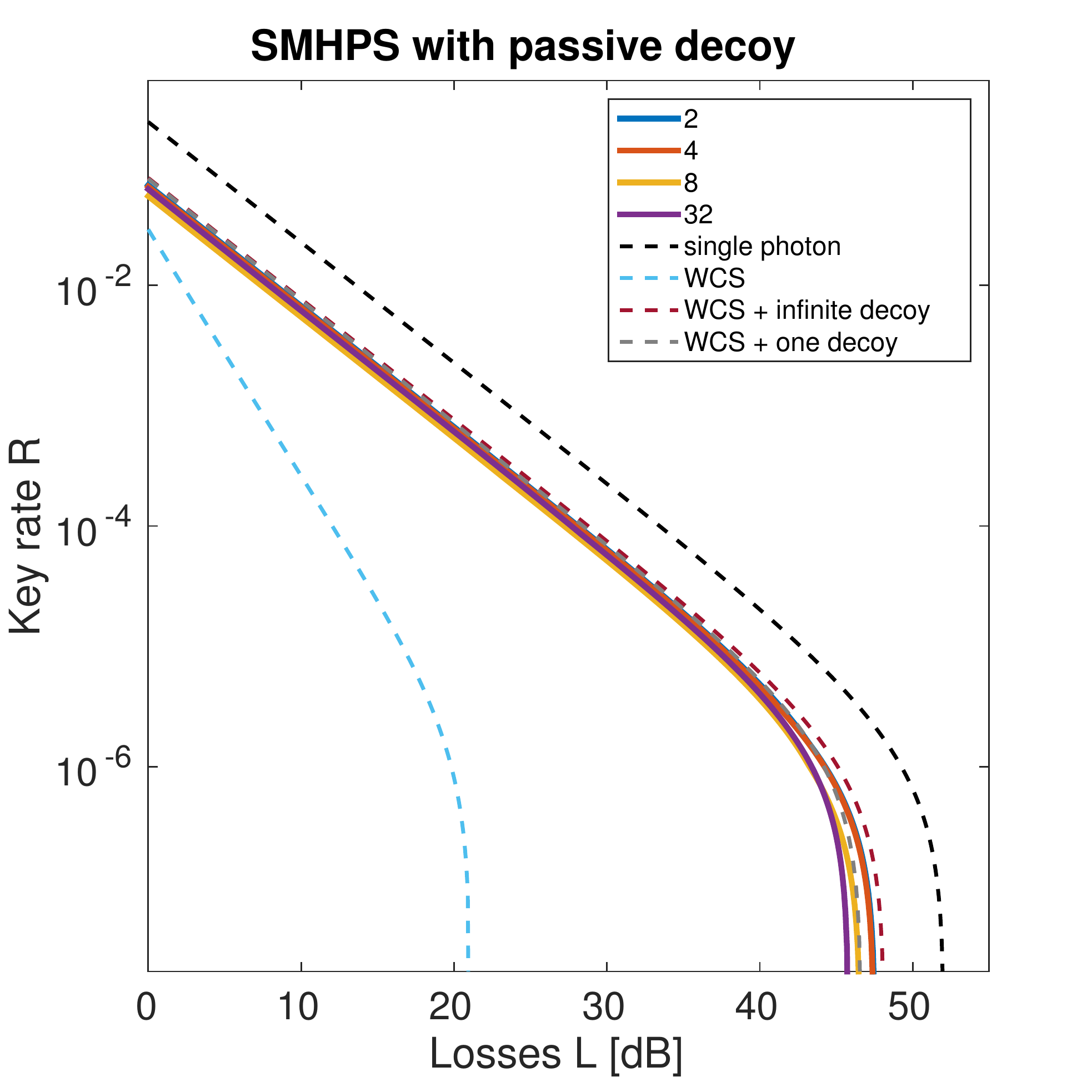}}
	{\includegraphics[width=0.45\linewidth]{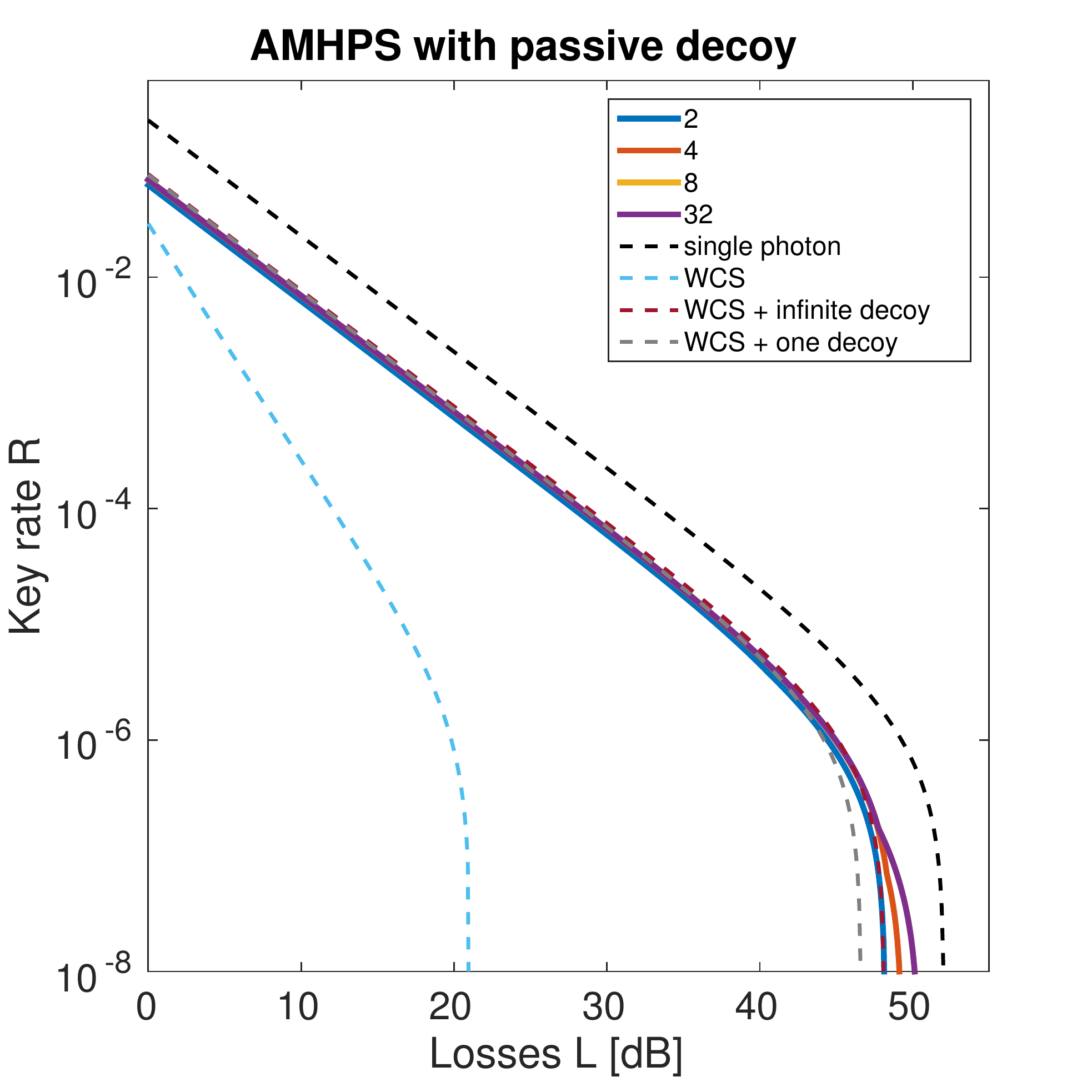}}
	\caption{Key rate for passive decoy state QKD for the SMHPS (left) and the AMHPS (right), with $\eta = 0.7$ and $\gamma = 0.5$.}
	\label{fig:passive}
\end{figure*}
The normalized mean number of generated pairs that maximizes the key rate is almost constant for both sources in the range $\sim 0.6 - 0.9$, with a steep fall in the regime where dark counts become important.
The rate of the SMHPS decreases as the number of crystals increases, thus further underlining the detrimental effect of the increased absorption in optical switches.
The AMHPS, on the other hand, does not show any improvement for more than 4 HS units, because the probability of triggering the rightmost HS units is negligible (see Appendix \ref{sec:source}). 

Since in the proposed implementation of passive decoy the bound on channel parameters is no longer optimal,  simulations have to take into account also this effect on the key rate.
Therefore, the key rate is calculated by inserting the bounds for $e_1$, $Y_0$ and $Y_1$ into (\ref{eq:passive_partial_rate}) and taking into account the fraction of pulses using each statistics with equation (\ref{eq:passive_rate}).
Simulation results are shown in Figure \ref{fig:passive}.
The use of just two different statistics gives a non-optimal parameter estimation, so a lower key rate than the schemes using active decoy.
To account for this, the rates in Figure \ref{fig:passive} are compared also with a similar, one decoy scheme implemented with attenuated laser pulses \cite{Ma2005}.
The worse bound on the parameters gives a worse estimation of the information leaked to Eve, thus requiring the sources to be pumped with lower intensity than in the case of active decoy (the normalized mean number of generated pairs oscillates, in this case, between $0.2$ and $0.3$).
The performance of the SMHPS is comparable, or slightly worse, to the one of decoy state with attenuated lasers, because of the already discussed detrimental effect on the source caused by optical switch attenuation.
On the other hand, the key rate of the AMHPS is always higher than the one obtained with the attenuated laser in the one decoy scheme and almost reaches the maximum tolerable loss level of the attenuated laser with decoy.
As in all previous schemes, the AMHPS shows no improvement after a certain threshold of HS units is reached  (in this case, $m = 4$). 

\begin{figure}
	\centering \includegraphics[width=0.9\linewidth]{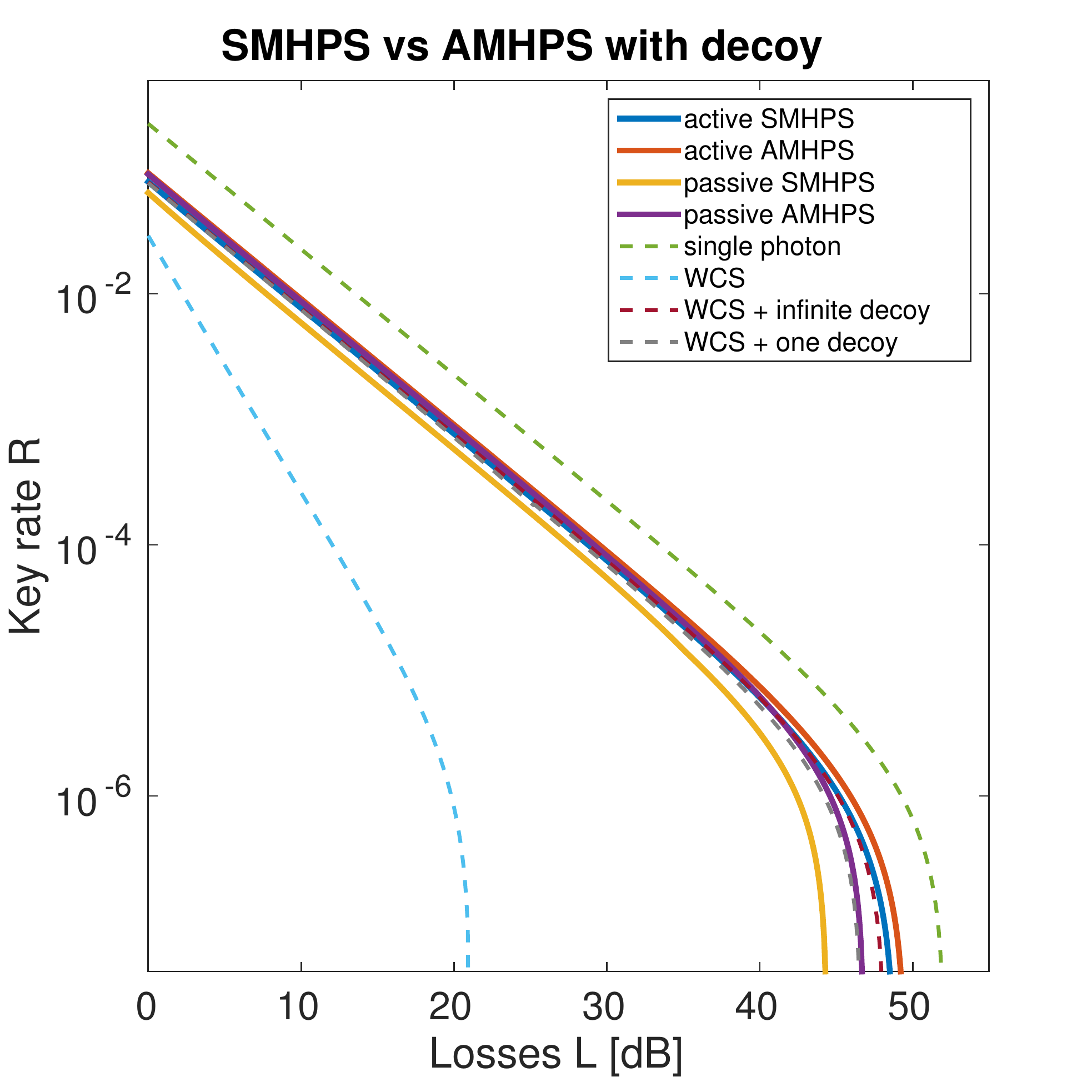}
	\caption{Key rate for the SMHPS and the AMHPS for both active and passive decoy with $m=8$, $\eta = 0.7$ and $\gamma = 0.5$.}
	\label{fig:decoy}
\end{figure}

The comparison of the two sources in the two different decoy schemes of Figure \ref{fig:decoy} directly shows the advantage of the AMHPS over the SMHPS.
Indeed, the asymmetric scheme performs better than the passive one in both active and passive decoy.
Furthermore, the key rate of the AMHPS in the passive scheme almost equals both the SMHPS and the attenuated laser with decoy, despite the worse parameter estimation caused by the use of just one decoy state.

\section{Conclusion}
The multiple-crystal heralded sources have shown better performance than the attenuated laser in all the studied cases, with the only exception of the SMHPS in the one decoy scheme, where the high absorption in optical switches, together with an imperfect parameter estimation, completely overrules the enhancement given by the multiple crystal configuration.
The AMHPS shows better scalability than the SMHPS, since in the latter the addition of HS units can degrade the performance while in the former it has, in the worst case scenario, no effect.

In addition to this, the key rate of the AMHPS is generally higher than for the SMHPS.
The asymmetric structure of the AMHPS, on the other hand, makes its implementation harder than the SMHPS, 
since each crystal much be fed a different pump intensity and the different HS units must be carefully synchronized.
Both architectures have shown a stronger dependence of the key rate on optical switch transmittance than on detector efficiency. 

The effect of multi-photon pulses, even if less important than for the attenuated laser, still limits the maximum tolerable loss level.
The use of the decoy state solves this problem also for multiple-crystal heralded sources.
The best results are given by active decoy, where a variable optical attenuator and a random number generator are used to change the output statistics.
The much more complex statistics of these sources, however, makes the calculations needed for the determination of the optimal decoy parameters impractical.
Therefore, the simplifying assumption of exact determination of the parameters has been adopted and the obtained results just give a superior limit on the attainable key rate.
The optimal decoy parameters must be calculated on a case by case basis, once the characteristics of the source have been chosen. 

These limitations might make it preferable the use of a passive decoy scheme, 
which can be easily implemented in both sources by exploiting the post-selection mechanism.
In this case, the optimization has taken into account also the inefficiencies of parameter estimation, thus the obtained results give the effective key rate attainable with each configuration and not just a superior limit. 

All the results here presented are subjected to the approximation of infinitely long key.
We leave finite key effects for future studies.
These effects are important especially for the schemes with decoy state, since each photon statistics requires a sufficiently high number of events.
In the passive decoy scheme the effect is still more important, 
since the need of considering also the relative frequency of the two photon statistics can drastically change the optimal source parameters. 

The recent advances in integrated photonics suggest an increasing role of multiple-crystal heralded sources.
If built into a single chip, they might be a valid alternative to lasers in quantum key distribution and other quantum information tasks requiring single photons to work properly.
Our analysis show the supremacy of the asymmetric scheme with respect to the symmetric one for QKD applications.

\begin{acknowledgments}
The Authors acknowledge the Strategic-Research-Project QUINTET of the Department of Information Engineering, University of Padova and the Strategic- Research-Project QUANTUMFUTURE of the Univer- sity of Padova.
MS acknowledges the Ministry of Research and the Center of Studies and Activities for Space (CISAS) ``Giuseppe Colombo'' for financial support.
\end{acknowledgments}

\appendix

\section{Multiple crystal heralded sources with post-selection}
\label{sec:source}
\begin{figure}
	\centering \includegraphics[width=.8\linewidth]{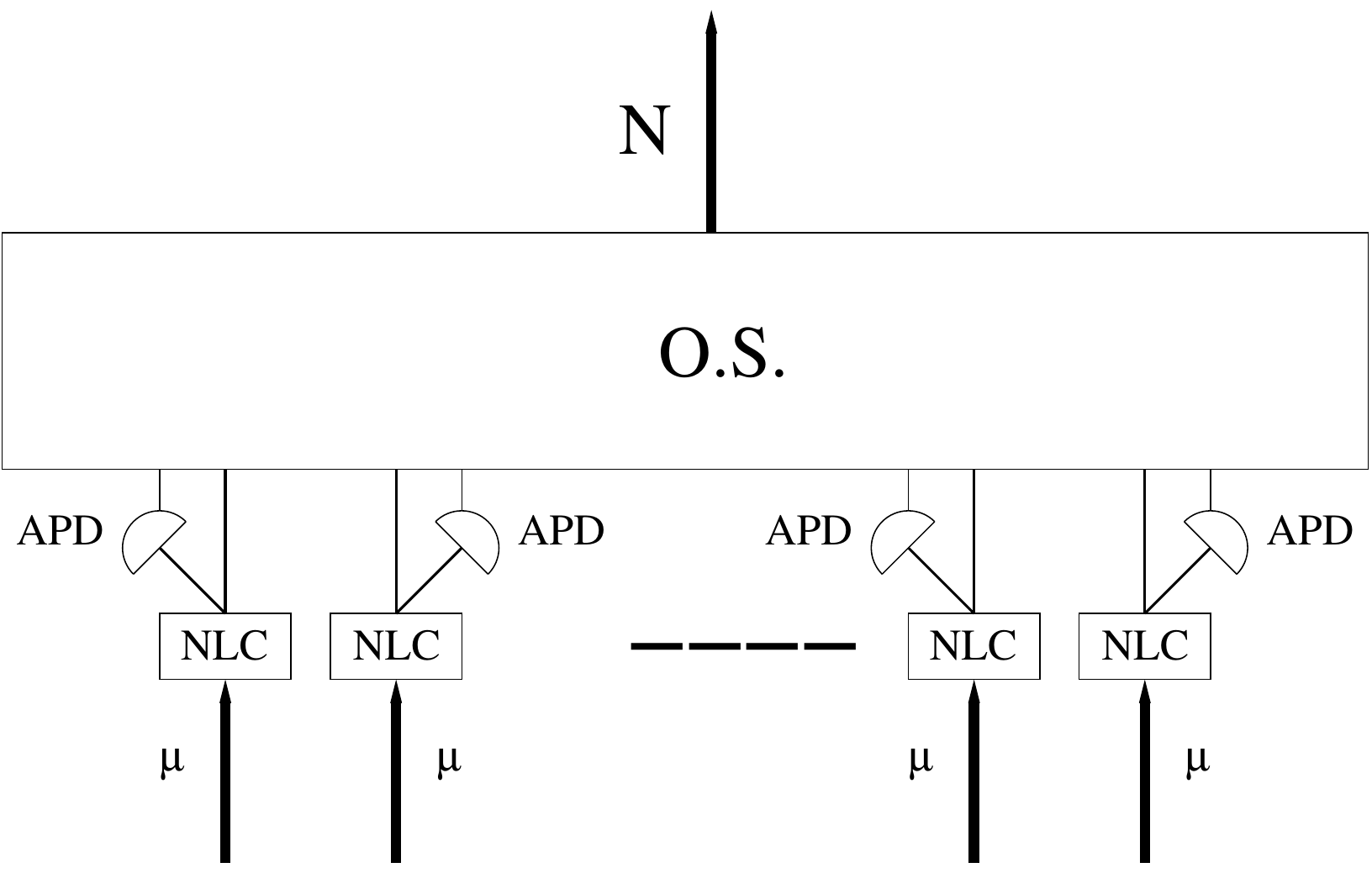}
	\caption{Schematic of the MHPS \cite{Migdall2002}. Each non-linear crystal (NLC) is fed with a pulse such that the mean number of generated pairs is $\mu$. The idler photon is fed into a detector (APD), while the signal one is routed through an optical switch (O.S.) to the output.}
	\label{fig:MHPS}
\end{figure}

The multiple crystal heralded source with post-selection (MHPS) consists on an array of $m$ HS units, simultaneously pumped with a laser pulse with intensity such that the mean number of generated pairs per pulse is $\mu$ \cite{Migdall2002}.
For each HS unit, labelled with index $i = 1..m$, the idler photon is used as a trigger for the signal one, which is injected into an optical switch, as shown in Figure \ref{fig:MHPS}.
In the ideal case of perfect detector and optical switch, a post-selection mechanism that selects one of the channels whose detector has fired gives the output statistics
\begin{equation}
	P_n^{M} (\mu; m) = \frac{\mu^n}{n!} e^{-\mu} \frac{1 - e^{-m \mu}}{1 - e^{-\mu}} (1 - \delta_n) + \delta_n e^{-m \mu},
\end{equation}
where $\delta_n$ is the Kronecker Delta ($\delta_0 = 1$ and $\delta_{n>0} = 0$) \cite{Mazzarella2013}. 

Taking into account real devices, the MHPS would face the low efficiency of $m$-to-1 optical switches.
This can be avoided by replacing the $m$-to-1 switch with a tree structure of 2-to-1 switches, giving the symmetric MHPS (SMHPS) \cite{Shapiro2007} shown in Figure \ref{fig:SMHPS}.

The structure of this scheme requires the number of HS units $m$ to be a power of 2.
The idler photon of each HS unit is fed into a detector with quantum efficiency $\eta$, while the signal photon is directed to a 2-to-1 optical switch of transmittance $\gamma$.
Since the photons produced by each HS unit pass $k = \log_2 m$ switches before reaching the output, the crystals are pumped with an intensity such that the mean number of generated pairs per pulse is $\mu/\gamma^k$.
The post-selection mechanism in each optical switch gives priority to the left HS unit and, in case no HS unit triggers, always outputs the left one \cite{Mazzarella2013}.
The photon statistics at the output is
\begin{multline}
	P_n^{S} (\mu; m, \eta, \gamma) = \frac{(1 - \eta) \mu e^{-(1-\eta) \mu}}{n!} e^{-\eta \mu \frac{2^k}{\gamma^k}} \\
	+ \frac{\mu^n e^{-\mu}}{n!} \frac{1 - (1-\eta)^n e^{-\eta ( \frac{1}{\gamma^k} - 1 ) \mu}}{1 - e^{-\eta \frac{\mu}{\gamma^k}}} (1 - e^{- \eta \mu \frac{2^k}{\gamma^k}} ).
\end{multline}
In the asymmetric MHPS (AMHPS), the $m$ HS units are arranged following the scheme shown in Figure \ref{fig:AMHPS} \cite{Mazzarella2013}.
Each HS unit, whose idler photon is fed into a detector of efficiency $\eta$, is pumped with an intensity such that the mean number number of generated pair per pulse is $\mu/\gamma^{k_i}$, with
\begin{equation}
	k_i = \left\{
		\begin{array}{l l}
			i & i \leq m-1 \\
			m-1 & i = m, \\
		\end{array}
		\right.
\end{equation}
in order to compensate the different number of traversed 2-to-1 optical switches.
The post-selection mechanism of each optical switch is the same as for the SMHPS, i.e. it gives priority to the left HS unit and, if none triggers, outputs the left one.
Differently from the SMHPS, this architecture requires delay lines to be introduced, in order to compensate the longer transmission time of the rightmost HS units \cite{Mazzarella2013}.
The statistics at the output is
\begin{multline}
	P_n^{A} (\mu; m, \eta, \gamma) = \frac{[(1-\eta) \mu] e^{-(1-\eta) \mu}}{n!} e^{-\eta \mu \frac{(2-\gamma)\gamma^{1-m} - 1}{1-\gamma}} \\
	+ \frac{\mu^n e^{-\mu}}{n!} \sum_{i=1}^{m} e^{-\eta \mu \frac{\gamma^{1-i}-1}{1-\gamma}} [ 1 - (1-\eta)^n e^{\eta \mu} e^{- \frac{\eta \mu}{\gamma^{k_i}}} ].
\end{multline}

\section{Parameter estimation in passive decoy state QKD}
\label{sec:passive}
The parameters $Y_0$, $Y_1$ and $e_1$, necessary in post-processing, are not directly measured during the key exchange session, but must be estimated from the experimental data $Q$ and $E$.
If Alice registers, for each pulse, whether at least one or no detector has clicked, a different gain and QBER for each case can be measured and these can be used for parameter estimation.
Since both statistics depend on the same $\mu$, parameter estimation can be included in the general optimization process, thus giving the real key rate and not just a superior limit. 

The probability that no detector clicks in a pulse is
\begin{equation}
	P^{nc} (\mu; m, \eta, \gamma) = e^{-\mu \eta \frac{2^k}{\gamma^k}}
\end{equation}
for the SMHPS and
\begin{equation}
	P^{nc} (\mu; m, \eta, \gamma) = e^{-\mu \eta \frac{(2-\gamma) \gamma^{1-m}-1}{1-\gamma}}
\end{equation}
for the AMHPS, where $k = \log_2 m$.
The statistics in the case of no click is the same for both sources (in both cases the first HS unit is routed to the output) and is
\begin{equation}
	P^{(nc)}_n = \frac{[\mu (1 - \eta)]^n}{n!} e^{-\mu (1-\eta)},
\end{equation}
while the statistics for the case of at least a detector click is
\begin{equation}
	P^{(c)}_n = \frac{\mu^n e^{-\mu}}{n!} [ 1 - (1-\eta)^n e^{-\eta \mu \left(\frac{1}{\gamma^k} - 1\right)} ] \frac{1}{1 - e^{\frac{-\mu \eta}{\gamma^k}}}
\end{equation}
for the SMHPS and
\begin{multline}
	P^{(c)}_n = \\
	\frac{\mu^n e^{-\mu}}{n!} \sum_{i=1}^{m} [ 1 - (1-\eta)^n e^{-\mu \eta \left( \frac{1}{\gamma^{k_i}} - 1 \right)} ] \frac{e^{-\mu \eta \frac{\gamma^{1-i}-1}{1-\gamma}}}{1 - e^{-\mu \eta \frac{(2-\gamma)\gamma^{1-m} - 1}{1-\gamma}}}
\end{multline}
for the AMHPS \cite{Mazzarella2013}. 

After the key exchange session, Alice tells Bob for which pulses at least one detector has clicked, so that they can estimate the gain and the QBER separately for the two cases.
From these values, referenced to as $\{Q^{c},E^{c},Q^{nc},E^{nc}\}$, and the known source statistics $P^{(c)}_n$ and $P^{(nc)}_n$, they can estimate the parameters of the channel using the method described in \cite{Curty2010}. 

The first parameter to be estimated is $Y_0$.
Its upper bound $Y^U_0$ can be calculated starting from the relations
\begin{align}
	Q^{c} E^{c} = \sum_{n=0}^{\infty} P^{(c)}_n Y_n e_n \geq P^{(c)}_0 Y_0 e_0 \\
	Q^{nc} E^{nc} = \sum_{n=0}^{\infty} P^{(nc)}_n Y_n e_n \geq P^{(nc)}_0 Y_0 e_0.
\end{align}
Since both inequalities must hold, the parameter $Y_0$ is upper bounded by
\begin{equation}
\label{Y0U}
	Y_0 \leq Y_0^U = \min \left\{ \frac{Q^{c} E^{c}}{P^{(c)}_0 e_0} , \frac{Q^{nc} E^{nc}}{P^{(nc)}_0 e_0} \right\}.
\end{equation}
Its lower bound $Y_0^L$ can be calculated from
\begin{multline}
	P^{(c)}_1 Q^{nc} - P^{(nc)}_1 Q^{c} = \sum_{n=0}^{\infty} (P_1^{(c)} P_n^{(nc)} - P_1^{(nc)} P_n^{(c)}) Y_n \\
	\leq (P_1^{(c)} P_0^{(nc)} - P_1^{(nc)} P_0^{(c)}) Y_0,
\end{multline}
that gives
\begin{equation}
\label{Y0L}
	Y_0 \geq Y_0^L = \max \left\{ \frac{P_1^{(c)} Q^{nc} - P_1^{(nc)} Q^{c}}{P_1^{(c)} P_0^{(nc)} - P_1^{(nc)} P_0^{(c)}} , 0 \right\},
\end{equation}
since, for both the SMHPS and the AMHPS
\begin{multline}
	P_1^{(c)} P_n^{(nc)} - P_1^{(nc)} P_n^{(c)} = \\
		A_{n,1} [ (1-\eta)^n - (1-\eta) ] \left\{
		\begin{array}{lr}
			\leq 0 & \text{for } n \geq 2 \\
			\geq 0 & \text{for } n = 0,
		\end{array}
	\right.															
\end{multline}
with $A_{n,1}$ a positive constant. 

The lower bound on the single photon yield $Y_0$ is calculated starting from
\begin{multline}
	P_2^{(c)} Q^{nc} - P_2^{(nc)} Q^{c} = \sum_{n=0}^{\infty} ( P_2^{(c)} P_n^{(nc)} - P_2^{(nc)} P_n^{(c)} ) Y_n \\
	\leq \sum_{n=0}^{1} (P_2^{(c)} P_n^{(nc)} - P_2^{(nc)} P_n^{(c)} ) Y_n,
\end{multline}
which leads to
\begin{multline}
\label{Y1L}
	Y_1 \geq Y_1^L = \\
	\max \left\{ \frac{P_2^{(c)} Q^{nc} - P_2^{(nc)} Q^{c} - (P_2^{(c)} P_0^{(nc)} - P_2^{(nc)} P_0^{(c)}) Y_0^U }{P_2^{(c)} P_1^{(nc)} - P_2^{(nc)} P_1^{(c)}} , 0 \right\},
\end{multline}
since
\begin{multline}
	P_2^{(c)} P_n^{(nc)} - P_2^{(nc)} P_n^{(c)} = \\
	A_{n,2} [ (1-\eta)^n - (1-\eta)^2 ] \left\{
		\begin{array}{lr}
			\leq 0 & \text{for } n \geq 2 \\
			\geq 0 & \text{for } n \leq 1
		\end{array}
	\right.															
\end{multline}
with $A_{n,2}$ positive. 

Similarly, the upper bound on $e_1$ is calculated from
\begin{multline}
	P_0^{(nc)} Q^{c} E^{c} - P_0^{(c)} Q^{nc} E^{nc} = \\
	\sum_{n=0}^{\infty} ( P_0^{(nc)} P_n^{(c)} - P_0^{(c)} P_n^{(nc)}) e_n Y_n \\
	\geq (P_0^{(nc)} P_1^{(c)} - P_0^{(c)} P_1^{(nc)}) e_1 Y_1,
\end{multline}
since
\begin{equation}
	P_0^{(nc)} P_n^{(c)} - P_0^{(c)} P_n^{(nc)} = A_{n,0} [ 1 - (1-\eta)^n ] \geq 0 
\end{equation}
for all $n$, and
\begin{align}
	Q^{c} E^{c} = \sum_{n=0}^{\infty} P_n^{(c)} Y_n e_n \geq P_0^{(c)} Y_0 e_0 + P_1^{(c)} Y_1 e_1, \\
	Q^{nc} E^{nc} = \sum_{n=0}^{\infty} P_n^{(nc)} Y_n e_n \geq P_0^{(nc)} Y_0 e_0 + P_1^{(nc)} Y_1 e_1,	
\end{align}
thus obtaining
\begin{multline}
\label{e1U}
	e_1 \leq e_1^U = \min \left\{ \frac{P_0^{(nc)} Q^{c} E^{c} - P_0^{(c)} Q^{nc} E^{nc}}{(P_0^{(nc)} P_1^{(c)} - P_0^{(c)} P_1^{(nc)}) Y_1^L} , \right. \\
	\left. \frac{Q^{c} E^{c} - P_0^{(c)} Y_0^L e_0}{P_1^{(c)} Y_1^L} , \frac{Q^{nc} E^{nc} - P_0^{(nc)} Y_0^L e_0}{P_1^{(nc)} Y_1^L} \right\}.
\end{multline}


\end{document}